\documentclass{article}[12pt]

\usepackage{epsfig}
\usepackage{graphicx}
\usepackage{color}
\usepackage{dcolumn}
\usepackage{bm}
\usepackage{amsmath}
\usepackage{amssymb}
\usepackage{amsxtra}
\usepackage{natbib}

\newtheorem{theorem}{Theorem}[section]
\newtheorem{definition}[theorem]{Definition}

\newtheorem{cor}[theorem]{Corollary}

\newtheorem{alg}[theorem]{Algorithm}

\renewcommand{\setminus}{-}

\def\bnprf{\noindent {\bf Proof} \ }
\def\edprf{$_{\Box}$}
\def\ubar#1{\underbar{$#1$}}

\def\C{\mathcal{C}^{\pm}}
\def\Cp{\mathcal{C}^{+}}
\def\Cm{\mathcal{C}^{-}}

\def\I{\mathcal{C}^{0}}

\def\CC{\vec{C}_{3}}
\def\L{L}

\def\de{arc}	
\def\DG{\vec{G}}

\def\DE{A}

\def\ngeq{\succeq}

\def\pgeq{\geq}

\newsavebox{\bsq}
\savebox{\bsq}(6,6){{\scriptsize $\bullet$}}
\newsavebox{\sq}
\savebox{\sq}(6,6){{\scriptsize $\circ$}}

\title{On uniform sampling simple directed graph realizations of degree sequences}

\author{M.~Drew LaMar\thanks{Department of Applied Science,
The College of William and Mary,
311 McGlothlin-Street Hall,
Williamsburg VA 23187 ({\tt mdlama@wm.edu}).}}

\begin{document}

\maketitle

\begin{abstract}
Choosing a uniformly sampled simple directed graph realization of a degree sequence has many applications, in particular in social networks where self-loops are commonly not allowed.  It has been shown in the past that one can perform a Markov chain arc-switching algorithm to sample a simple directed graph uniformly by performing two types of switches: a 2-switch and a directed 3-cycle reorientation.  This paper discusses under what circumstances a directed 3-cycle reorientation is required.  In particular, the class of degree sequences where this is required is a subclass of the directed 3-cycle anchored degree sequences.  An important implication of this result is a reduced Markov chain algorithm that uses only 2-switches.
\end{abstract}

\section{Introduction}

Markov chain Monte Carlo algorithms have been used successfully to uniformly sample realizations of both undirected and directed degree sequences \cite{Cobb:2003p2229,MR1662523}.  The algorithms use a sequence of moves from a move-set to go from one realization to another.  This results in a random walk on a {\it meta-graph}, where each vertex corresponds to a realization and the edges connecting these vertices correspond to moves from the move-set.  If the meta-graph is connected with appropriate probability weights for the edges (see \cite{Cobb:2003p2229}), we will be guaranteed a uniformly sampled realization with the fixed degree sequence.

To sample {\it simple} directed realizations (i.e. no self-loops or multi-arcs), there are two types of moves in our move-set \cite{MR1662523}: a 2-switch and the reorientation of a directed 3-cycle $\CC$, where $\CC$ has vertex set $\{v_{1},v_{2},v_{3}\}$ and arc set $\{(v_{1},v_{2}),(v_{2},v_{3}),(v_{3},v_{1})\}$.  A 2-switch is given by
\begin{center}
\medskip
\includegraphics[height=0.7in]{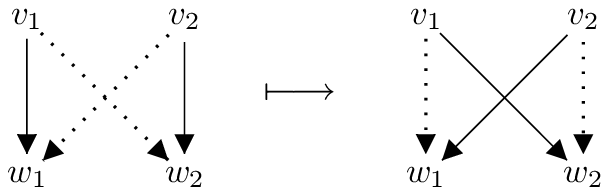}
\medskip
\end{center}
where dotted lines denote no arcs, with the $\CC$ reorientation given by
\begin{center}
\medskip
\includegraphics[height=0.7in]{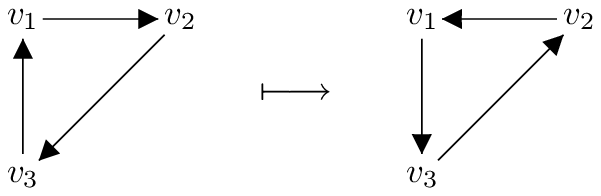}.
\medskip
\end{center}
$\CC$ reorientations can lead to a much larger mixing time in certain circumstances.  In this paper we identify the cases where $\CC$ reorientations are necessary, give a degree-sequence characterization of these cases, and show that we can reduce our move-set to only 2-switches.

Recently, Berger and M{\"u}ller-Hannemann posted a paper with similar results.
In \cite{Berger:2009p11038}, they rediscover the result by Rao et al.~\cite{MR1662523} proving connectivity of the meta-graph using 2-switches and $\CC$ reorientations.  They also implement a Monte Carlo algorithm similar to Rao et al.~\cite{MR1662523} which uniformly samples simple realizations from a directed degree sequence, including mixing time calculations as well.  They show, as we do, that the special cases where $\CC$ reorientations are required are precisely the subset of $\CC$-anchored digraphs which we call $\CC^{*}$-anchored.  Our paper differs in that our proof is substantially shorter (built upon the structural characterization of $\CC$-anchored digraphs found in \cite{LaMar:2009p9967}) and uses the degree sequence characterization of the $\CC$-anchored digraphs to identify the $\CC$-anchors, as opposed to a more computationally intensive algorithm that requires the knowledge of all induced 3-cycles of a given realization.  Using the degree sequence characterization is much faster (linear in the number of vertices) and allows us to use the more efficient 2-switch random walk.  We both, however, show that the meta-graph consists of $2^{k}$ isomorphic subgraphs, where $k$ is the number of anchored 3-cycles.

\section{Notation}

All directed graphs in this article will be simple, i.e. with no self-loops or multi-arcs.  We consider integer-pair sequences $d=\{(d_{i}^{+},d_{i}^{-})\}_{i=1}^{N}$ and say $d$ is {\it digraphic} if there exists a digraph (i.e. directed graph) with degree sequence $d$, denoting the set of digraph realizations of $d$ by $R(d)$.  All integer-pair sequences are assumed to be digraphic (otherwise $R(d) = \emptyset$), and thus $d^{+}$ and $d^{-}$ will denote the out-degree and in-degree sequences of $d$, respectively.

We denote directed graphs by $\DG$, with $V(\DG)$ the vertex set and $\DE(\DG)$ the {\de } set.  We will drop the reference to $\DG$ when the digraph is understood through the notation $\DG = (V,\DE)$, for example.  An {\de } between vertices $a$ and $b$ will be denoted by $(a,b)$, with the orientation given by the ordering.  

Given a digraph $\DG = (V,\DE)$ and vertex sets $X,Y\subset V$, we define the subgraph $\DG[X,Y] = (X\cup Y, \DE[X,Y])$, where $\DE[X,Y] = \{(x,y)\in \DE : x \in X \ \mbox{and} \ y \in Y\}$.  When $X=Y$, we have the usual definition of an induced subgraph and will denote this by $\DG[X]$.

We will use the vertex labeling notation $v_{i}$ in place of $L^{-1}(i)$, where $L$ is a bijective labeling function $\L: V \longrightarrow \{1,\ldots,|V|\}$ going from vertices to coordinates of the degree sequence.

\section{Result}

Given a degree sequence $d$, we define the {meta-graph} $\Omega_{d} = (\mathcal{V},\mathcal{E})$, where $\mathcal{V}$ is in one-to-one correspondence with $R(d)$.  We will denote $V_{\DG} \in \mathcal{V}$ to be the vertex corresponding to $\DG \in R(d)$.  There are two types of edges $\mathcal{E} = \mathcal{E}_{2}\cup \mathcal{E}_{3}$: $(V_{\DG},V_{\DG^{\prime}}) \in \mathcal{E}_{2}$ if there is a 2-switch between them.  Similarly, $(V_{\DG},V_{\DG^{\prime}}) \in \mathcal{E}_{3}$ if there is a $\CC$ reorientation connecting them.  We have the following result:

\begin{theorem}[Rao et al.~\cite{MR1662523}]
The meta-graph $\Omega_{d}$ is connected.
\label{thm:conn}
\end{theorem}

We can define a Markov chain random walk on $\Omega_{d}$ by an appropriate choice of probability weights for each edge in $\mathcal{E}$.  There are many choices for the weights, but for simplicity I will give as an example probability weights induced by a particularly simple random walk algorithm (see \cite{Roberts:2000p6832, Cobb:2003p2229}).  Given a realization $\DG^{(n)}\in R(d)$, with probability $p$ attempt a 2-switch and with probability $1-p$ a $\CC$ reorientation.  For a 2-switch, choose four vertices without replacement and, if possible, perform a 2-switch to arrive at $\DG^{(n+1)}$.  Otherwise, do nothing, i.e. $\DG^{(n+1)} = \DG^{(n)}$.  Similarly, for a $\CC$ reorientation choose three vertices without replacement and, if possible, perform a $\CC$ reorientation to arrive at $\DG^{(n+1)}$.  Otherwise, do nothing.  The resulting probabilities for this Markov chain are given by
\[P_{ij}=\begin{cases} 
p/{N\choose{4}} & \text{if $(V_{\DG_{i}}, V_{\DG_{j}}) \in \mathcal{E}_{2}$},\\ 
(1-p)/{N\choose{3}} & \text{if $(V_{\DG_{i}}, V_{\DG_{j}}) \in \mathcal{E}_{3}$}.
\end{cases}\]
By doing nothing with failed move attempts, we impose self-loops at each realization such that $\sum_{j=1}^{|R(d)|}P_{ij} = 1$.  By Theorem \ref{thm:conn}, this Markov chain is irreducible, and it is easily seen to be symmetric and aperiodic.  Thus, there is a unique limiting distribution which by symmetry must be the uniform distribution.

It is mentioned in \cite{Roberts:2000p6832, MR1662523} that in most situations one need only use 2-switches, and thus we can choose $p$ to be close to 1.  The difficulty with this is there are degree sequences where this will lead to very long mixing times, due to the rare cases where there is not a path with edges in $\mathcal{E}_{2}$ connecting two realizations.  The rarity of these cases is also unknown, and so there is no way to know how close to 1 one should choose $p$.  What are the structure of these degree sequences, and can we identify them?  It turns out we can identify them: they are a subset of what are known as {\bf $\bm \CC$-anchored} degree sequences, as defined below.

\begin{definition}
We call a degree sequence $d$ {\bf $\bm \CC$-anchored} if it is forcibly $\CC$-digraphic and there exists a nonempty set of coordinates $J$, called a {\bf $\bm \CC$-anchor set}, such that for every coordinate $i \in J$ and every $\DG \in R(d)$, there is an induced subgraph $\vec{C}^{\prime}\subseteq \DG$ with $\vec{C}^{\prime}\cong \CC$ and $v_{i} \in V(\vec{C}^{\prime})$.  All realizations $G \in R(d)$ are also called {\bf $\bm \CC$-anchored digraphs}.
\label{def:anchor}
\end{definition}

The structural characterization of $\CC$-anchored digraphs was given in \cite{LaMar:2009p9967} by a digraph decomposition using $M$-partitions.  An $M$-partition of a digraph $\DG$ is a partition of the vertex-set $V(\DG)$ into $k$ disjoint classes $\{X_{1},\ldots,X_{k}\}$, where the {\de } constraints within and between classes are given by a symmetric $k\times k$ matrix $M$ with elements in $\{0,1,*\}$ (see \cite{Feder:2003p7761}).  $M_{ii}$ equals $0$ or $1$ when $X_{i}$ is an independent set or clique, respectively, and is set to $*$ when $\DG[X_{i}]$ is an arbitrary subgraph.  Similarly, for $i\neq j$, $M_{ij}$ equal to $0$, $1$, or $*$ corresponds to $\DG[X_{i},X_{j}]$ having no {\de}s from $X_{i}$ to $X_{j}$, all {\de}s from $X_{i}$ to $X_{j}$, and no constraints on {\de}s from $X_{i}$ to $X_{j}$, respectively. 

\begin{figure}[t]
\centering
\includegraphics[width=5in]{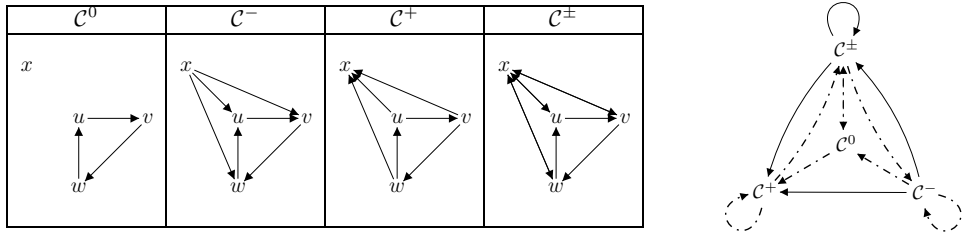}
\caption{{\bf Left}: The 4 vertex classes $\{\I,\Cm,\Cp,\C\}$ of $\CC^{*}$-anchored digraphs defined by how a vertex $x$ in each class connects to a directed 3-cycle $\CC$ with vertex set $\{u,v,w\}$.  {\bf Right}: Diagram showing the relations within and between the 4 possible vertex classes.  Solid and dashed-dotted arrows denote forced and allowable {\de}s, respectively, while the absence of an arrow denotes no {\de}s.  $\C$ is a clique, $\I$ an independent set, while $\DG[\Cm]$ and $\DG[\Cp]$ are arbitrary subgraphs.}
\label{fig:3rules}
\end{figure}

The subset of $\CC$-anchored digraphs that are the focus of this paper are called $\CC^{*}$-anchored and are realizations of $\CC$-anchored degree sequences such that $|J| = 3K$, where $K$ is a positive integer, with $\DG[\{v_{j_{3n+1}},v_{j_{3n+2}},v_{j_{3n+3}}\}] \simeq \CC$ for all realizations $\DG \in R(d)$, $0 \leq n \leq K-1$.  Note that $K$ denotes the number of {\it anchored} 3-cycles, i.e. those vertices that induce a directed 3-cycle for all realizations.  $\CC^{*}$-anchored digraphs have a structural characterization given by the following theorem (see Fig.~\ref{fig:3rules} for a pictorial representation):

\begin{theorem}[LaMar~\cite{LaMar:2009p9967}]
The digraph $\DG=(V,\DE)$ is a $\CC^{*}$-anchored digraph if and only if there is a $C\subset V$ such that $\DG[C]\cong \CC$ and an $M$-partition of $\DG[V\setminus C]$ with vertex classes given by $\{\I,\Cm,\Cp,\C\}$, where each class defines how its elements relate to $C$ as follows:
\begin{eqnarray*}
	\I \ & \equiv & \ \{ x \in V\setminus C \ : \ (x,C)\cup(C,x) \subset \DE^{C} \} \\
	\Cm \ & \equiv & \ \{ x \in V\setminus C \ : \ (x,C) \subset \DE \ \mbox{and} \ (C,x) \subset \DE^{C} \} \\
	\Cp \ & \equiv & \ \{ x \in V\setminus C \ : \ (C,x) \subset \DE \ \mbox{and} \ (x,C) \subset \DE^{C} \} \\
	\C \ & \equiv & \ \{ x \in V\setminus C \ : \ (x,C)\cup(C,x) \subset \DE \}
\end{eqnarray*}
and matrix $M$ given by
\[
\bordermatrix{
    & \C & \Cm & \Cp & \I \cr
\C  &  1 &  *  &  1  &  * \cr
\Cm &  1 &  *  &  1  &  * \cr
\Cp &  * &  0  &  *  &  0 \cr
\I  &  * &  0  &  *  &  0
}.
\]
\label{def:gT}
\end{theorem}

If we define the meta-graph $\Omega_{d}^{\prime} = (\mathcal{V},\mathcal{E}_{2})$, then the following is the main theorem of this paper.

\begin{theorem}
$\Omega_{d}^{\prime}$ is disconnected if and only if $d$ is $\CC^{*}$-anchored.
\label{thm:disconnected}
\end{theorem}

\bnprf
Let $d$ be $\CC^{*}$-anchored and $\DG \in R(d)$.  It should be clear that $\Omega_{d}^{\prime}$ is disconnected, since every realization has a directed 3-cycle through the same three vertices, and thus no 2-switches will connect realizations with the opposite orientations for that 3-cycle without first removing the 3-cycle.

\begin{figure}[t!]
\includegraphics{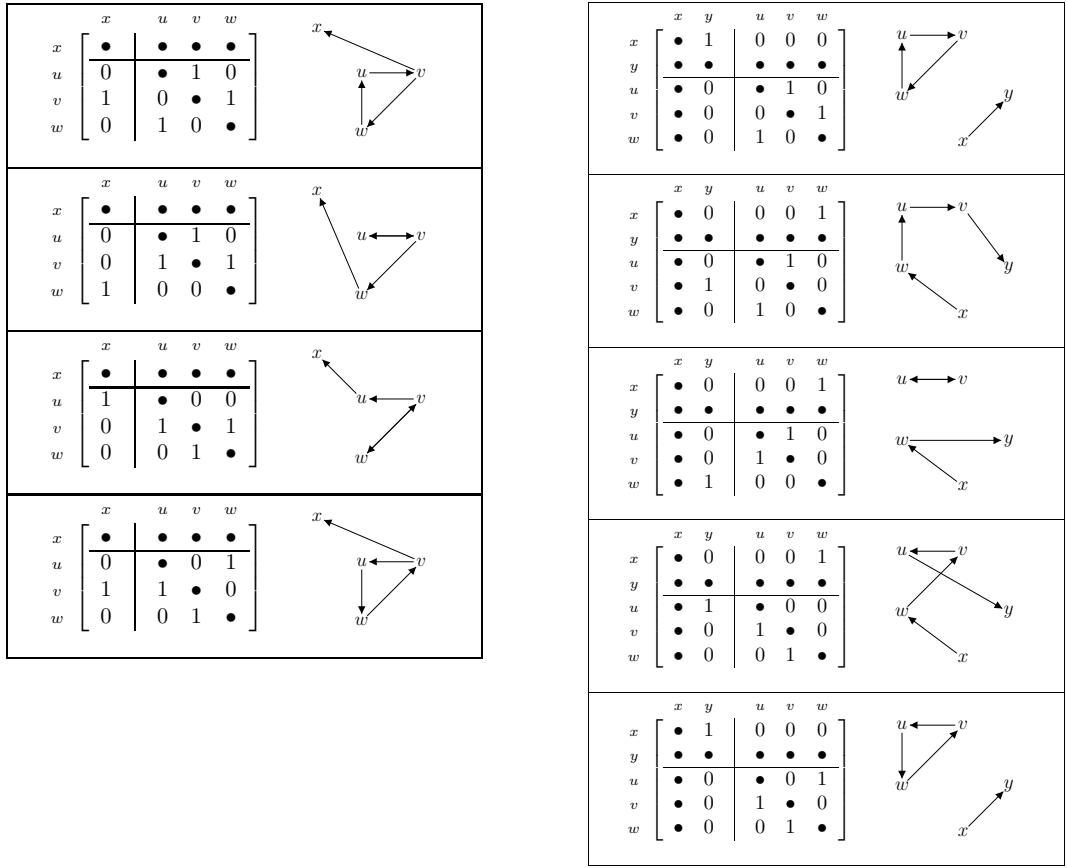}
\caption{Both left and right panels show a series of 2-switches used to reorient a directed 3-cycle $\CC$ (see Theorem~\ref{thm:disconnected}).  The solid dots denote entries of the adjacency matrix which are not used in the series of moves.}
\label{fig:picprf}
\end{figure}

Suppose $d$ is a degree sequence with $\Omega_{d}^{\prime}$ disconnected.  By Theorem \ref{thm:conn}, there must be a realization $\DG \in R(d)$ with an oriented directed 3-cycle $C$ such that $V_{\DG}$ is in one connected component of $\Omega_{d}^{\prime}$ and $V_{\DG^{\prime}}$ is in another connected component, where $\DG^{\prime}$ is found from $\DG$ by reorienting $C$.  Let $C = \{u,v,w\}$ with $\{(u,v),(v,w),(w,u)\} \subset A$.  We want to show that for any vertex $x \in V-C$, $x$ must be in one of the vertex classes $\I$, $\Cm$, $\Cp$ or $\C$.  We will show that $(C,x) \subset A$ and/or $A^{C}$ (by symmetry, we will also have $(x,C) \subset A$ and/or $A^{C}$).  Suppose there is only one of the three arcs, and without loss of generality choose $(v,x)\in A$, with $\{(u,x),(w,x)\} \subset A^{C}$ (the case with two existing arcs follows by considering the graph complement).  The left panel in Fig.~\ref{fig:picprf} shows that we can perform a series of 2-switches to reorient the 3-cycle, which contradicts $\DG$ and $\DG^{\prime}$ being in two separate connected components of $\Omega_{d}^{\prime}$.  Thus, we must have $(C,x) \subset A$ and/or $A^{C}$, showing $x \in \I$, $\Cm$, $\Cp$ or $\C$.

Now we must show that the connections between the vertex classes are given by the $M$-matrix in Theorem~\ref{def:gT}.  Let $x \in \I \cup \Cp$ and $y \in \I \cup \Cm$, and suppose that $(x,y) \in A$.  In the right panel of Fig.~\ref{fig:picprf}, we see again that there is a series of 2-switches which reorients the 3-cycle, showing $(x,y)\notin A$.  By considering the graph complement, we can prove $(x,y) \in A$ for $x \in \C \cup \Cm$ and $y \in \C \cup \Cp$.  This shows $d$ is $\CC^{*}$-anchored, thereby completing the proof.

\edprf

For every anchored 3-cycle $C$, there are two isomorphic copies of connected components of $\Omega^{\prime}_{d}$ corresponding to each orientation of $C$.  In general, $\Omega_{d}$ thus has the following form.

\begin{cor}
\[\Omega_{d} \simeq \Omega_{d}[\mathcal{V}(G_{2})]\times\left(\times_{i=1}^{k}K_{2}\right),\]
where $G_{2}$ is one connected component of $\Omega^{\prime}_{d}$ and $k$ denotes the number of anchored 3-cycles.
\end{cor}

The real power of this result is the knowledge that if we know where the anchored 3-cycles are, then we can simply choose an orientation for each anchored 3-cycle uniformly at random, and then perform a random walk on the graph $\Omega_{d}^{\prime}$.  This will be an efficient procedure if the identification of the anchored 3-cycles can be done without too much work.  It was shown in \cite{LaMar:2009p9967} that $\CC^{*}$-anchored digraphs have not only a structural characterization as given in Theorem~\ref{def:gT} but also a degree-sequence characterization.  In other words, we can identify the anchored 3-cycles using a simple procedure on the degree sequence itself.  To this end, we start with some definitions.

Given an integer sequence $a$, define the {\it corrected conjugate sequence} $a^{\prime\prime}$ by
\[
a_{k}^{\prime\prime} = |I_{k}| + |J_{k}|,
\]
where
\begin{eqnarray*}
I_{k} & = & \{i \ | \ i < k \ \ \mathrm{and} \ \ a_{i} \geq k-1\}, \\
J_{k} & = & \{i \ | \ i > k \ \ \mathrm{and} \ \ a_{i} \geq k\}.
\end{eqnarray*}

\begin{definition}
A degree sequence $d = \{(d^{+}_{i}, d^{-}_{i})\}_{i=1}^{N}$ is non-increasing relative to the {\bf positive lexicographical ordering} if and only if $d^{+}_{i} \geq d^{+}_{i+1}$, with $d^{-}_{i} \geq d^{-}_{i+1}$ when $d^{+}_{i} = d^{+}_{i+1}$.  In this case, we will call $d$ {\bf positively ordered} and denote the ordering by $d_{i} \pgeq d_{i+1}$.  We say $d$ is non-increasing relative to the {\bf negative lexicographical ordering} by giving preference to the second coordinate, calling $d$ in this case {\bf negatively ordered} and denoting the ordering by $d_{i} \ngeq d_{i+1}$.
\end{definition}

For a given degree sequence $d = \{(d^{+}_{i},d^{-}_{i})\}_{i=1}^{N}$, define the sequences $\bar{d} = \{(\bar{d}^{+}_{i}, \bar{d}^{-}_{i})\}_{i=1}^{N}$ and $\ubar{d} = \{(\ubar{d}^{+}_{i}, \ubar{d}^{-}_{i})\}_{i=1}^{N}$ to be the positive and negative orderings of $d$, respectively.

For a degree sequence $d$, define the {\it slack} sequences $\bar{s}$ and $\ubar{s}$ by
\begin{eqnarray*}
\bar{s}_{l} & = & \sum_{i=1}^{l}[\bar{d}^{-}]_{i}^{\prime\prime}-\sum_{i=1}^{l}\bar{d}^{+}_{i} \ \ \mbox{with} \ \ \bar{s}_{0} \equiv 0, \\
\ubar{s}_{k} & = & \sum_{i=1}^{k}[\ubar{d}^{+}]_{i}^{\prime\prime}-\sum_{i=1}^{k}\ubar{d}^{-}_{i} \ \ \mbox{with} \ \ \ubar{s}_{0} \equiv 0.
\end{eqnarray*}

\begin{theorem}[LaMar~\cite{LaMar:2009p9967}]
The degree sequence $d = \{(d^{+}_{i},d^{-}_{i})\}_{i=1}^{N}$ is $\CC^{*}$-anchored if and only if there are coordinates $\{j_{1}, j_{2}, j_{3}\}$ and an integer-pair $(k,l) \geq (1,1)$ such that
\begin{equation}
d_{j_1} = d_{j_2} = d_{j_3} = (k,l)
\label{equ:gS1}
\end{equation}
with
\begin{equation}
(d_{j_{1}}, d_{j_{2}}, d_{j_{3}}) = (\bar{d}_{l}, \bar{d}_{l+1}, \bar{d}_{l+2}) = (\ubar{d}_{k}, \ubar{d}_{k+1}, \ubar{d}_{k+2})
\label{equ:gS2}
\end{equation}
and the slack sequences satisfying
\begin{equation}
(0, 1, 1, 0) = (\bar{s}_{l-1}, \bar{s}_{l}, \bar{s}_{l+1}, \bar{s}_{l+2}) = (\ubar{s}_{k-1}, \ubar{s}_{k}, \ubar{s}_{k+1}, \ubar{s}_{k+2}).
\label{equ:gS3}
\end{equation}
In this case, $\{v_{j_{1}},v_{j_{2}},v_{j_{3}}\}$ induces an anchored 3-cycle.
\label{thm:anchor}
\end{theorem}

\begin{alg}
To achieve a uniformly sampled simple realization of a degree sequence $d$, we check if $d$ is $\CC^{*}$-anchored and identify the anchored 3-cycles using Theorem~\ref{thm:anchor}, randomly assign an orientation to each anchored 3-cycle with equal probability (which effectively chooses a connected component of $\Omega_{d}^{\prime}$), and then perform a random walk on this component of $\Omega_{d}^{\prime}$ using only 2-switches.
\end{alg}


\section{Conclusion}

We have shown that the degree sequences that require both types of move sets, i.e. 2-switches and directed 3-cycle reorientations, are the $\CC^{*}$-anchored degree sequences whose degree sequences have been characterized in \cite{LaMar:2009p9967}.  This characterization allows for a fast algorithm that identifies the $\CC$-anchor sets, leading to a Monte Carlo algorithm involving only 2-switches.

This is the second instance where $\CC$-anchored degree sequences have been found to be the special cases in algorithms involving directed graphs (for the first case, see \cite{LaMar:2009p9967}).  It is interesting to see where else this structure may be important.

\section*{Acknowledgments}
The author thanks Gregory Smith and Sarah Day for helpful discussions.  This work was funded by the postdoctoral and undergraduate biological sciences education program grant awarded to the College of William and Mary by the Howard Hughes Medical Institute.

\bibliography{mcmc01}

\begin{thebibliography}{1}

\bibitem{Berger:2009p11038}
Annabell Berger and Matthias M{\"u}ller-Hannemann.
\newblock Uniform sampling of undirected and directed graphs with a fixed
  degree sequence.
\newblock {\em arXiv}, 0912.0685v1, Dec 2009.

\bibitem{Cobb:2003p2229}
George~W Cobb and Yung-Pin Chen.
\newblock An application of markov chain monte carlo to community ecology.
\newblock {\em The American Mathematical Monthly}, 110(4):265--288, 2003.

\bibitem{Feder:2003p7761}
T~Feder, P~Hell, S~Klein, and R~Motwani.
\newblock List partitions.
\newblock {\em SIAM J. Discrete Math.}, 16(3):449--478, Jan 2003.

\bibitem{LaMar:2009p9967}
M.~Drew LaMar.
\newblock Algorithms for realizing degree sequences of directed graphs.
\newblock {\em arXiv}, 0906.0343v1, June 2009.

\bibitem{MR1662523}
A~Rao, Rabindranath Jana, and Suraj Bandyopadhyay.
\newblock A markov chain monte carlo method for generating random
  (0,1)-matrices with given marginals.
\newblock {\em Sankhya Ser. A}, 58(2):225--242, 1996.

\bibitem{Roberts:2000p6832}
John Roberts.
\newblock Simple methods for simulating sociomatrices with given marginal
  totals.
\newblock {\em Social Networks}, 22(3):273--283, 2000.

\end{thebibliography}
\bibliographystyle{plain}

\end{document}